\documentclass[12pt,a4,english]{article}
\usepackage[latin9]{inputenc}
\usepackage{geometry}
\usepackage{verbatim}
\usepackage{amsmath}
\usepackage{amsfonts}
\usepackage{amsthm}
\usepackage{amssymb}
\usepackage{mathtools}
\usepackage{graphicx}
\usepackage{longtable}
\usepackage{setspace}
\usepackage{comment}
\usepackage{xcolor}
\usepackage{upgreek}

\usepackage{tikz,xcolor}

\definecolor{lime}{HTML}{A6CE39}
\DeclareRobustCommand{\orcidicon}{%
	\begin{tikzpicture}
	\draw[lime, fill=lime] (0,0) 
	circle [radius=0.16] 
	node[white] {{\fontfamily{qag}\selectfont \tiny ID}};
	\draw[white, fill=white] (-0.0625,0.095) 
	circle [radius=0.007];
	\end{tikzpicture}
	\hspace{-2mm}
}
\foreach \x in {A, ..., Z}{%
	\expandafter\xdef\csname orcid\x\endcsname{\noexpand\href{https://orcid.org/\csname orcidauthor\x\endcsname}{\noexpand\orcidicon}}
}

\usepackage[round,sort,authoryear]{natbib}
\usepackage[colorlinks=true,linkcolor=blue, citecolor=blue, urlcolor = blue]{hyperref}

\usepackage{titlesec}
\usepackage{babel}
\usepackage[toc,page]{appendix}
\usepackage{booktabs,caption}

\usepackage[labelfont=bf,textfont=md]{caption}
\usepackage[labelsep=space]{caption}

\usepackage{lmodern}
\usepackage[T1]{fontenc}

\DeclarePairedDelimiter{\floor}{\lfloor}{\rfloor}

\usepackage{bibentry}
\nobibliography*

\usepackage{fancyhdr}
\numberwithin{equation}{section}

\titleformat*{\section}{\large \bfseries}
\titleformat*{\subsection}{\normalsize \bfseries}
\titleformat*{\subsubsection}{\small \bfseries}

\newif\ifshow 
\showfalse 

\ifshow
  \includecomment{wrap}
\else
  \excludecomment{wrap} 
\fi

\theoremstyle{definition}

\newtheorem{example}{Example}

\begin{document}
\pagenumbering{roman}

\title{\Large \textbf{Unified Inference for Dynamic Quantile Predictive Regression}\thanks{\textbf{Article history:} First draft - September 26, 2023, Second draft - November 10, 2023. I am grateful to Jose Olmo, Tassos Magdalinos and Jean-Yves Pitarakis from the School of Economic, Social and Political Sciences, University of Southampton for helpful conversations. Moreover, I am also grateful to Jayeeta Bhattacharya, Pietro Spini and Byunghoon Kang for stimulating discussions on relevant aspects to this paper. Financial support from the Research Council of Finland (grant 347986) is gratefully acknowledged.} \\ }

\author{\textbf{Christis Katsouris}\thanks{Postdoctoral Researcher, Economics Division, Faculty of Social Sciences, University of Helsinki, P.O. Box 17, Arkadiankatu 7 Helsinki, FI-00014 Finland. E-mail: \textcolor{blue}{\texttt{christis.katsouris@helsinki.fi}} } \\ \textit{University of Helsinki}\\ \\ \textcolor{blue}{First Version:} September 26, 2023\\ \textcolor{blue}{This Version:}  November 10, 2023}

\date{}

\maketitle

\begin{abstract}
\vspace*{-1 em}
This paper develops unified asymptotic theory for dynamic quantile predictive regression models which holds regardless of the presence of intercept without a priori knowledge on the persistence properties of regressors. Our framework establishes the asymptotic distribution of the proposed test statistics based on two commonly used estimation methods for dealing with nonstationarity in predictive regressions. A finite-sample analysis of the proposed testing procedure will be examined via Monte Carlo simulations to compare the performance of these two competing estimation methods. Using a stock return dataset with monthly frequency we can also study the quantile predictability puzzle when the model has a dynamic specification. 
\\

\textbf{Keywords:} persistence; local-to-unity; dynamic regression; quantile predictability. 
\\

\textbf{JEL} Classification: C12, C22
\end{abstract}

\newpage 

\setcounter{page}{1}
\pagenumbering{arabic}

\section{Introduction}

Modelling financial and macroeconomic data commonly exhibit features such as structural breaks, nonstationarity in the form of near unit roots, stochastic trends or cointegration dynamics.  Motivated by the growing interest in estimating the risk measures of $\mathsf{VaR}$ and $\mathsf{CoVaR}$ using the local-to-unity parametrization which allows to capture the persistence properties of regressors (see,  \cite{katsouris2021optimal}), we focus on the dynamic quantile predictive regression. Limit results for predictive regressions with conditional mean (see, \cite{chen2013uniform}, \cite{kostakis2015Robust}, \cite{kasparis2015nonparametric}, \cite{li2017uniform}, \cite{yang2023unified}, \cite{harvey2023improved})  and conditional quantile functional forms (see, \cite{lee2016predictive}, \cite{fan2019predictive} and \cite{liu2023unified}), have already been extensively examined under various econometric conditions. Similarly unified inference in a dynamic predictive regression with a conditional mean functional form has been proposed by \cite{yang2021unified}. To the best of our knowledge the relevant asymptotic theory for conditional quantile functional forms in dynamic quantile predictive regressions is novel.    

Due the unknown persistence properties of regressors, imposing a local-to-unity specification\footnote{The limit theory for moderate deviations from a unit root in autoregressive processes have been examined by   \cite{giraitis2006uniform}, \cite{jansson2006optimal}, \cite{Phillips2007limit}, \cite{aue2007limit} and \cite{huang2014limit} among many others.} allows to capture various nonstationarity regimes; although with the challenge that one has to ensure asymptotic theory is nuisance parameter-free. Both the empirical likelihood approach (see, \cite{zhu2014predictive} and \cite{cai2014testing}) as well as the IVX filtration method (see, \cite{PhillipsMagdal2009econometric} and \cite{kostakis2015Robust})  ensure that statistical inference is uniform regardless of whether the predicting variable is stationary or nonstationary or has an infinite variance. Inference for a multivariate nonstationary regressor or even within a high-dimensional setting still involves challenges in both cases but is beyond the scope of this paper.  

The predictive regression model has been widely used for investigating the predictability of asset returns. Applications include estimation and inference methodologies for predictive regression model based on a parametric (see, \cite{kostakis2015Robust}) or nonparametric (see, \cite{kasparis2015nonparametric}) conditional mean functional form. The current literature mainly covers linear conditional mean autoregressive and predictive regression models as well as dynamic conditional mean predictive regression models. Thus, when one is interested for testing the presence of quantile predictability, related studies have considered the implementation of the model based on the conditional quantile functional form. Related studies include \cite{maynard2023inference}, \cite{lee2016predictive}, \cite{fan2019predictive}, \cite{katsouris2022asymptotic},  \cite{cai2022new} and \cite{liu2023unified}.  Furthermore, the inclusion of model intercepts in both the predictive regression model as well as the autoregressive model either based on a conditional mean or on a conditional quantile functional forms imposes additional challenges when developing inference techniques for the $\beta$ parameter based on the usual likelihood ratio test or Wald-type statistics;  especially when the autoregressive coefficient is nearly-integrated (see, \cite{romano2001subsampling}, \cite{liu2019asymptotic},  \cite{wang2022testing}, \cite{yang2023unified}). 

Several studies in the literature employ the likelihood-based approach to investigate the properties and asymptotic theory of corresponding estimators and test statistics for nonstationary autoregressive processes and predictive regression models. These likelihood approaches include the weighted least squares, the empirical likelihood methodology as well as the profile likelihood approach.  In particular, \cite{zhu2014predictive},  \cite{liu2019empirical, liu2019unified} and \cite{yang2021unified} consider an application of the empirical likelihood approach which induces uniform inference across the different degrees of persistence and confidence regions for the coefficients of the linear predictive regression. Moreover, the framework of \cite{chen2013uniform} proposes a uniform inference approach (see, also \cite{chen2009bias}\footnote{Related studies include \cite{chen2009restricted}, \cite{chen2012restricted} as well as \cite{zhao2000restricted}.}) which is valid regardless of the persistence of regressors using the weighted least squares approximate restricted likelihood (WLSRL).  \cite{liu2019unified} propose predictability tests that correspond to  testing the null hypothesis of no statistical significance for the model coefficients of a predictive regression model with  additional covariates the difference of the predicting variable. These predictability tests are based on the unified empirical likelihood methodology that rejects the null hypothesis $H_0: \beta_1 = \beta_2 = 0$ or testing the hypothesis $H_0: \beta_1 = 0$ and $H_0: \beta_2 = 0$. Under the presence of known nonzero model intercept in the predictive regression model using weighted score equations and the sample-splitting methodology it can be proved that nuisance parameter-free inference holds regardless of the persistence properties of regressors. 


Both methodologies are fundamentally aiming to develop asymptotic theory which is valid to the presence of model intercept in the predictive regression model as it is well-known that nonstandard limit distribution results appear when $\left\{ X_t \right\}_{t=1}^n$ is nonstationary. The empirical likelihood method takes the first difference of the paired data such that $\left\{ Y_{t+1} - Y_t \right\}_{t=1}^n$ and $\left\{ X_{t+1} - X_t \right\}_{t=1}^n$, but a new issue arises as the corresponding error sequence $\left\{ U_{t+1} - U_t \right\}_{t=1}^n$ violates the independence assumption. Therefore, to fix this issue a split sample methodology is applied and the first difference transformation is applied based on a larger lag (i.e., the value of $m = \floor{n/2}$) which ensures $m-$dependence holds (i.e., approximate or asymptotic independence as $m \to \infty$), and lastly the empirical likelihood estimation is applied. The asymptotic theory for the empirical likelihood confidence interval for $\beta_0$ is then proved to follow standard distribution results (i.e., converges to a $\chi^2-$distribution). The IVX filtration is an endogenous instrumentation procedure which is used as an alternative to the OLS estimator for the unknown model coefficient $\beta_0$. In particular, the IVX instruments are constructed with the main consideration of obtaining less persistence regressors than the original series $\left\{ X_t \right\}$ without having to consider a first-difference transformation based on a large value of the lag as well as the application of sample splitting to ensure that the distribution of the innovations of the predictive regression have independent realizations. This methodology implies that asymptotic theory for the unknown model coefficient follows a mixed gaussian distribution and the corresponding Wald test follow a $\chi^2-$distribution.  

Uniform inference for unit root moderate deviations (based on the local-to-unity parametrization) holds in both cases. The empirical likelihood plus weighted least squares induces uniform inference regardless of persistence properties or the presence of an intercept in the autoregressive equation. Likewise, robustness to the abstract degree of persistence holds for the IVX estimator and uniform inference can be established based on the mixed gaussianity property (see, \cite{kostakis2015Robust}). Recently \cite{Magdalinos2022uniform} show that the IVX-instrumentation is indeed robust to the presence of an intercept under general autoregressive dependence. Similar properties hold  for conditional quantile functional forms; \cite{lee2016predictive} develops a framework for the implementation of the IVX instrumentation in quantile predictive regressions, which is found to have standard asymptotic theory while \cite{liu2023unified}  proved that their approach is indeed robust to both the presence of intercept in the autoregressive equation and nonstationarity of the $\left\{ X_t \right\}$ process. 



The rest of the paper is organized as follows. Section \ref{Section2}, presents the model estimation procedure and main assumptions of our framework. Section \ref{Section3}, examines the finite-sample performance of the proposed unified tests using a Monte Carlo simulation study and Section \ref{Section4} illustrates the implementation of the method using real data. Section \ref{Section5} summarizes the main conclusions. Proofs of the main results can be found in the Appendix of the paper.


\section{Econometric Model of Dynamic Quantile Predictive Regression}
\label{Section2}

\subsection{Classical Predictive Regression}

Let $y_t$ be a realization of the random variable $\left\{ Y_t \right\}_{t=1}^n$, which corresponds to a univariate predictant (such as stock return) that is assumed to be stationary, and $x_t$ a realization of the random variable $\left\{ X_t \right\}_{t=1}^n$, which corresponds to   be a predictive variable which is assumed to be possibly nonstationary. The pair $\left\{ ( Y_t, X_{t-1} ) \right\}_{t=1}^n$  is assumed to follow the predictive regression structure: 
\begin{align}
Y_t &= \alpha_n + \beta_n X_{t-1} + U_t,  \ \ \ \text{for} \ \  1 \leq t \leq n \\
X_t &= \mu_n + \rho_n X_{t-1} + V_t 
\end{align}
where $\rho_n = \left( 1 + \frac{ c }{ k_n } \right)$ with $k_n = n^{\gamma}$, such that $\gamma = 0$,  $\gamma = 1$ or  $\gamma \in (0,1)$. Moreover, assume that the initial condition $X_0$ is constant and $\left( U_t, V_t \right)$ are independently and identically distributed errors with mean zero and finite variance. Consider the vector $W_t = \left( U_t, V_t \right)^{\prime}$ such that $W_t$ is a strictly stationary martingale difference sequence. Then, the partial sums process for the sequence $W_t$ satisfy the invariance principle as proposed by \cite{Phillips1987time, Phillips1987towards} (see, \cite{Phillips1992asymptotics}) 
\begin{align}
\frac{1}{ \sqrt{n} } \sum_{t=1}^{\floor{ns} } W_t \Rightarrow \mathbf{B}_{z} (s) := \big[ B_u (s) , B_v (s) \big]^{\top} \equiv \text{BM} \left( \mathbf{\Sigma}_{ww} \right), \ \ \mathbf{\Sigma}_{ww} 
:= 
\begin{bmatrix}
\Sigma_{uu} & \Sigma_{uv} \\
\Sigma_{uv} & \Sigma_{vv} 
\end{bmatrix}
\end{align}
where $\mathbf{B}_{w}(s)$ is a vector Brownian motion with a positive-definite matrix  $\mathbf{\Sigma}_{ww}$ such that $\Sigma_{uu} >  0$, $\Sigma_{uu} > 0$ and $\Sigma_{vv} > 0$.

\subsubsection{Variable-Augmented Predictive Regression}

The econometric specification for the predictive regression model with variable-augmentation is: 
\begin{align}
Y_t &= \alpha_n + \gamma_n Y_{t-1} + \beta_n X_{t-1} + U_t,  \ \ \ \text{for} \ \  1 \leq t \leq n \\
X_t &= \mu_n + \rho_n X_{t-1} + V_t 
\end{align}
In terms of statistical inference, based on the above conditional mean functional form, the unknown parameters of interest are $\theta_n = ( \beta_n, \gamma_n )$. A relevant hypothesis testing formulation is the joint existence of the lagged predicted variables and predictability, which implies that under the null hypothesis it holds that $\mathcal{H}_0: \left\{ \beta_n = 0 \right\} \ \bigcap \ \left\{ \gamma_n = 0 \right\}$. Unified tests which are robust to the unknown degree of persistence, imply that uniform inference regardless of regressors being stationary or nearly intergated as well as in the case that the autoregressive equation is specified with or without an intercept holds. In particular, \cite{li2017uniform} assumes that $\gamma$ is unknown and tests either $\mathcal{H}_0: \gamma = 0$ or $\mathcal{H}_0: \beta = 0$ without requiring that $\gamma = 0$. A simplified approach is the case that one assumes that $\gamma = 0$ and formulates the null hypothesis $\mathcal{H}_0: \beta = 0$ (see, \cite{demetrescu2014enhancing}). However, is more economically meaningful to avoid imposing the assumption that the model parameter $\gamma = 0$ a priori. Then, the main interest is to construct a robust inference method for testing the null hypothesis that $\mathcal{H}_0: \gamma = 0$ and $\mathcal{H}_0: \beta = 0$,  regardless of the following cases:
\begin{itemize}
\item $\left\{ X_t \right\}$ is stationary: $| \rho_0 | < 1$ 

\item $\left\{ X_t \right\}$ is nearly-integrated: $\rho_0 =  \left( 1 - \frac{c}{n} \right)$ for some $c \neq 0$. 

\item $\left\{ X_t \right\}$ is unit root: $\rho_0 = 1$.    
\end{itemize}
Under the assumption that $\alpha_0 = 0$ then, the least squares estimator for $( \gamma, \beta )^{\top}$ can be obtained by solving the following score equations: 
\begin{align}
\frac{ \partial S_t }{ \partial \gamma } &=  \sum_{t=1}^n \big( Y_t - \alpha_0 - \gamma Y_{t-1} - \beta X_{t-1} \big) Y_{t-1} \equiv 0
\\
\frac{ \partial S_t }{ \partial \beta } &=  \sum_{t=1}^n \big( Y_t - \alpha_0 - \gamma Y_{t-1} - \beta X_{t-1} \big) X_{t-1} = 0
\end{align}
Therefore, we can construct an inference procedure about the model parameters $\beta$ and $\gamma$ by directly applying the empirical likelihood method based on estimating equations. However, the asymptotic theory of the empirical likelihood function is nonstandard when $\left\{ X_t \right\}$ is nearly-integrated.

\newpage

\subsubsection{Empirical Likelihood Estimation Approach}

The empirical likelihood approach is defined as below: 
\begin{align}
L_n (\beta) = \mathsf{sup} \left\{ \prod_{t=1}^n (n p_t): p_1 \geq 0,..., p_n \geq 0, \sum_{t=1}^n p+t = 1, \sum_{t=1}^n p_t Z_t ( \beta )     \right\}    
\end{align}
where $Z_t (\beta) = \big( Y_t - \beta X_{t-1} \big) X_{t-1} \big/ \sqrt{ 1 + X^2_{t-1} }$. Then, based on the Lagrange multiplier optimization method it follows that 
\begin{align}
\ell_n (\beta) := - 2 \mathsf{log} L_n (\beta) = 2 \sum_{t=1}^n \mathsf{log} \big\{ 1 + \lambda Z_t (\beta) \big\}    
\end{align}
Then, taking the first-order-condition for some $\lambda = \lambda (\beta)$ satisfies
\begin{align}
\sum_{t=1}^n \frac{ Z_t (\beta) }{ 1 + \lambda Z_t (\beta) } \equiv 0.    
\end{align}

\subsection{Dynamic Quantile Predictive Regression Model}

Recall that the quantile predictive regression model has the advantage that the innovation sequences of the predictive regression based on a conditional quantile functional form, denote with $\psi_{ \uptau } ( u_{t} ( \uptau ) )$ are only contemporaneously correlated with the innovations of unit root sequences. Our main contribution to the literature is to propose a unified inference procedure for Dynamic Quantile Predictive Regressions (DQPRs), which can be considered as further extension to the frameworks proposed by \cite{liu2023unified} and \cite{yang2021unified}.

Consider the dynamic quantile predictive regression model: 
\begin{align}
y_t &= \alpha_{\tau} + \beta_{\tau} x_{t-1} + \gamma_{\tau} y_{t-1} + \epsilon_{t \tau}, \ \  \epsilon_{t \tau} = u_t - \sigma_t Q_{\eta} (\tau) 
\\
x_t &= \mu + \rho x_{t-1} + u_t
\end{align}
Moreover, consider the following quantile moment conditions as expressed below
\begin{align}
\mathbb{E} \big[ \big( \boldsymbol{1} \left\{ y_t \leq \beta x_{t-1}  \right\} - \tau  \big) \cdot x_{t-1} \big] = 0    
\end{align}
Following the framework proposed by \cite{whang2006smoothed} we replace the indicator function which is not differentiable in the whole of the support of the model parameter $\beta$ with a \textit{smooth function}.

\newpage

Define the following two score functions:
\begin{align}
Z^{\alpha}_{t, \tau} ( \beta, \gamma ) &:= \psi_{\tau} \big( Y_t - \alpha_0 - \beta X_{t-1} - \gamma Y_{t-1} \big) \equiv 0
\\
Z^{\beta}_{t, \tau} ( \beta, \gamma ) &:= \psi_{\tau} \big( Y_t - \alpha_0 - \beta X_{t-1} - \gamma Y_{t-1} \big) w_t  \equiv 0, \ \ w_t =  \frac{ X_{t-1} }{ \sqrt{ 1 + X^2_{t-1} }  }
\\
Z^{\gamma}_{t, \tau} ( \beta, \gamma ) &:= \psi_{\tau}  \big( Y_t - \alpha_0 - \beta X_{t-1} - \gamma Y_{t-1} \big) \widetilde{w}_t \equiv 0, \ \ \widetilde{w}_t = ( Y_{t-1} - \beta X_{t-1} ).
\end{align}
In particular the random variable $X_{t-1}$ is replaced by a weighted random variable $\displaystyle \frac{ X_{t-1} }{  \phi ( | X_{t-1} | )  }$ where the weight function $\phi (t)$ satisfies the limit $\frac{ t }{  \phi (t) } \to 1$ as $t \to \infty$. Thus a simple choice which satisfies these requirements is $\phi(t) = \sqrt{1 + t^2}$ which implies the weight $w(t) = \frac{t}{ \sqrt{1 + t^2} }$ that is indeed a bounded continuous function. Notice that self-weighted M-estimators are proved in the literature to have normal asymptotic theory.  Furthermore, the purpose of weight $w_t$ is to downweight the leverage points in $X_t$ ensuring this way the finiteness property of covariance matrices under the presence of heavy-tailed errors. Furthermore, the weight is added in the second estimating equation to ensure that $\frac{1}{n} \sum_{t=1}^n \frac{ X^2_{t-1} }{ 1 + X^2_{t-1} }$ converges in probability to a constant regardless the nonstationarity of the regressor series $\left\{ X_t \right\}$.  

Then we define with 
\begin{align}
\boldsymbol{Z}_t ( \beta, \gamma ) = 
\begin{pmatrix}
Z^{\alpha}_{t, \tau} ( \beta, \gamma )
\\
Z^{\beta}_{t, \tau} ( \beta, \gamma )
\\
Z^{\gamma}_{t, \tau} ( \beta, \gamma )
\end{pmatrix}
\end{align}
Therefore, the empirical likelihood approach satisfies the following equation: 
\begin{align}
\sum_{t=1}^n \frac{ \boldsymbol{Z}_t  ( \beta, \gamma )  }{ 1 + \boldsymbol{\lambda}^{\top} \boldsymbol{Z}_t  ( \beta, \gamma )  } \equiv 0.   
\end{align}
Furthermore define the following matrices $\boldsymbol{V}_n := \mathbb{E} \big[ \boldsymbol{Z}_t  ( \beta, \gamma ) \boldsymbol{Z}^{\prime}_t  ( \beta, \gamma )  \big]$. Similarly defining the standardized Lagrangian multiplies $\tilde{\lambda} = \boldsymbol{V}_n^{1/2} \lambda$ and the standardized score function $\boldsymbol{W}_t ( \beta, \gamma ) = \boldsymbol{V}_n^{1/2} \boldsymbol{Z}_t (\beta, \gamma )$ we have the following log-likelihood function. 
\begin{align}
\ell_n ( \beta, \gamma ) = 2 \sum_{t=1}^n \mathsf{log}  \big( 1 + \tilde{\lambda} \cdot \boldsymbol{W}_t ( \beta, \gamma ) \big).    
\end{align}
Then, developing a Taylor expansion for $\tilde{\lambda}$ for $\tilde{\lambda}$ and $\ell_h ( \beta_0 )$ implies that around the FOC: 
\begin{align*}
0 &= \frac{1}{n} \sum_{t=1}^n \frac{ \boldsymbol{W}_t( \beta, \gamma) }{ 1 + \tilde{\lambda} \cdot \boldsymbol{W}_t (\beta, \gamma ) }    
\\
&=
\frac{1}{n} \sum_{t=1}^n \boldsymbol{W}_t (\beta, \gamma ) - \left( \frac{1}{n} \sum_{t=1}^n \boldsymbol{W}_t (\beta, \gamma ) \boldsymbol{W}_t (\beta, \gamma )^{\prime} \right) \lambda + \frac{1}{n} \sum_{t=1}^n  \big(  \lambda^{\prime} \boldsymbol{W}_t (\beta, \gamma ) \big)^2  \boldsymbol{W}_t (\beta, \gamma ) - .... 
\end{align*}

\newpage

Therefore, it can be proved that:
\begin{align*}
\ell_h ( \beta_0 ) 
&= 
\left[ \frac{1}{\sqrt{n}} \sum_{t=1}^n \big( \boldsymbol{Z}_t ( \beta, \gamma ) - \mathbb{E} \boldsymbol{Z}_t ( \beta, \gamma ) \big) + \mathsf{g}(n) \mathbb{E} \boldsymbol{Z}_t ( \beta, \gamma ) \right]^{\prime} \times  \boldsymbol{V}_n^{-1} 
\\
& \times 
\left[ \frac{1}{\sqrt{n}} \sum_{t=1}^n \big( \boldsymbol{Z}_t ( \beta, \gamma ) - \mathbb{E} \boldsymbol{Z}_t ( \beta, \gamma ) \big) + \mathsf{g}(n) \mathbb{E} \boldsymbol{Z}_t ( \beta, \gamma ) \right]
\end{align*}
where $\mathsf{g}(n) = n^{1/2}$. 
In the stationary case when $| \rho | < 1$ we have that as $n \to \infty$
\begin{align}
\boldsymbol{D}_{1,m}
\begin{pmatrix}
\sqrt{m} \left( \hat{\alpha_{\tau}} - \alpha_{\tau} \right)   
\\
\sqrt{m} \left( \hat{\beta_{\tau}} - \beta_{\tau} \right)   
\end{pmatrix}
\overset{d}{\to} \mathcal{N} \big( 0, \boldsymbol{\Sigma}_1 \big).
\end{align}
\begin{align}
\boldsymbol{D}_{1,m} = f_{\epsilon}(0)
\begin{pmatrix}
1 &  \displaystyle \frac{1}{m} \sum_{t=1}^m x_{t-1}
\\
\displaystyle \frac{1}{m} \sum_{t= m + 1}^{2m} w_t & \displaystyle \frac{1}{m} \sum_{t=m+1}^{2m} w_t x_{t-1}
\end{pmatrix}
\ \ \ \text{and} \ \ \ 
\boldsymbol{\Sigma}_1 = \tau (1 - \tau) 
\begin{pmatrix}
1  & 0
\\
0  & \displaystyle \mathbb{E} \left(  \frac{ x_1^2 }{ 1 + x_1^2 } \right)
\end{pmatrix}
\end{align}
\begin{align}
A_{1,m} \times \sqrt{m} \left( \hat{\beta}_{\tau} - \beta_{\tau}  \right) \overset{d}{\to} \mathcal{N} \left( 0, \boldsymbol{\gamma}_1^{\top} \boldsymbol{\Sigma}_1 \boldsymbol{\gamma}_1 \right)    
\end{align}
where 
\begin{align*}
\boldsymbol{\gamma}_1 = 
\begin{pmatrix}
\displaystyle - \mathbb{E} \left( \frac{ x_t^2 }{ \sqrt{ 1 + x_t^2 } } \right) 
\\
1
\end{pmatrix}
, \   
A_{1,m} = \left\{ \frac{1}{m} \sum_{t=m+1}^{2m} x_{t-1} w_t - \left( \frac{1}{m} \sum_{t=m+1}^{2m} w_t \right) \left( \frac{1}{m} \sum_{t=m+1}^{2m} x_{t-1} \right) \right\} f_{\epsilon}(0).
\end{align*}
Denote with $\boldsymbol{\Lambda}_t = ( 1, x_t)^{\top}$ such that 
\begin{align}
\boldsymbol{Z}_m( \boldsymbol{v} ) =
\begin{pmatrix}
\displaystyle \frac{1}{ \sqrt{m} } \sum_{t=1}^m \psi_{ \tau } \big( \varepsilon_t - \boldsymbol{v}^{\top} \mathbb{M}_m^{-1} \boldsymbol{\Lambda}_{t-1} \big)
\\
\displaystyle \frac{1}{ \sqrt{m} } \sum_{t=m+1}^{2m} \psi_{ \tau } \big( \varepsilon_t - \boldsymbol{v}^{\top} \mathbb{M}_m^{-1}\boldsymbol{\Lambda}_{t-1} \big) w_t
\\
\displaystyle \frac{1}{ \sqrt{m} } \sum_{t=m+1}^{2m} \psi_{ \tau } \big( \varepsilon_t - \boldsymbol{v}^{\top} \mathbb{M}_m^{-1}\boldsymbol{\Lambda}_{t-1} \big) \widetilde{w}_t
\end{pmatrix}
\end{align}
Then, for any given $\boldsymbol{v} \in \mathbb{R}^2$ we have that 
\begin{align}
\boldsymbol{Z}_m( \boldsymbol{v} )
= 
\begin{pmatrix}
\displaystyle \frac{1}{ \sqrt{m} } \sum_{t=1}^m \psi_{ \tau } ( \varepsilon_t )
\\
\displaystyle \frac{1}{ \sqrt{m} } \sum_{t=m+1}^{2m} \psi_{ \tau } ( \varepsilon_t ) w_t
\\
\displaystyle \frac{1}{ \sqrt{m} } \sum_{t=m+1}^{2m} \psi_{ \tau } ( \varepsilon_t ) \widetilde{w}_t
\end{pmatrix}
- \boldsymbol{D}_m \boldsymbol{v} + o_p(1), \ \text{as} \ n \to \infty,
\end{align}    
\begin{proof}
Since it holds that $\psi_{\tau} = \tau - \boldsymbol{1} \left\{ u \leq 0   \right\}$, we have that
\begin{align*}
\boldsymbol{Z}_m( \boldsymbol{v} ) - 
\begin{pmatrix}
\displaystyle \frac{1}{ \sqrt{m} } \sum_{t=1}^m \psi_{ \tau } ( \varepsilon_t )
\\
\displaystyle \frac{1}{ \sqrt{m} } \sum_{t=m+1}^{2m} \psi_{ \tau } ( \varepsilon_t ) w_t
\\
\displaystyle \frac{1}{ \sqrt{m} } \sum_{t=m+1}^{2m} \psi_{ \tau } ( \varepsilon_t ) \widetilde{w}_t
\end{pmatrix}
&=
- 
\begin{pmatrix}
\displaystyle \frac{1}{ \sqrt{m} } \sum_{t=1}^m \left[ \boldsymbol{1} \big\{ \varepsilon_t \leq \boldsymbol{v}^{\top} \mathbb{M}_m^{-1} \boldsymbol{\Lambda}_{t-1} \big\} - \boldsymbol{1} \big\{ \varepsilon_t \leq 0  \big\} \right]
\\
\displaystyle \frac{1}{ \sqrt{m} } \sum_{t=m+1}^{2m}\left[ \boldsymbol{1} \big\{ \varepsilon_t \leq \boldsymbol{v}^{\top} \mathbb{M}_m^{-1} \boldsymbol{\Lambda}_{t-1} \big\} - \boldsymbol{1} \big\{ \varepsilon_t \leq 0  \big\} \right] \cdot  w_t
\\
\displaystyle \frac{1}{ \sqrt{m} } \sum_{t=m+1}^{2m}\left[ \boldsymbol{1} \big\{ \varepsilon_t \leq \boldsymbol{v}^{\top} \mathbb{M}_m^{-1} \boldsymbol{\Lambda}_{t-1} \big\} - \boldsymbol{1} \big\{ \varepsilon_t \leq 0  \big\} \right] \cdot  \widetilde{w}_t    
\end{pmatrix}
\\
&=
- 
\begin{pmatrix}
\displaystyle \frac{1}{ \sqrt{m} } \sum_{t=1}^m \mathbb{E} \big[ \xi_{1,t} | \mathcal{F}_{t-1} \big]
\\
\displaystyle \frac{1}{ \sqrt{m} }  \sum_{t=m+1}^{2m} \mathbb{E} \big[ \xi_{2,t} | \mathcal{F}_{t-1} \big]
\\
\displaystyle \frac{1}{ \sqrt{m} }  \sum_{t=m+1}^{2m} \mathbb{E} \big[ \xi_{3,t} | \mathcal{F}_{t-1} \big]
\end{pmatrix}
- 
\begin{pmatrix}
\displaystyle \frac{1}{ \sqrt{m} } \sum_{t=1}^m \xi_{1,t} - \mathbb{E} \big[ \xi_{1,t} | \mathcal{F}_{t-1} \big]  
\\
\displaystyle \frac{1}{ \sqrt{m} }  \sum_{t=m+1}^{2m} \xi_{2,t} - \mathbb{E} \big[ \xi_{2,t} | \mathcal{F}_{t-1} \big]  
\\
\displaystyle \frac{1}{ \sqrt{m} }  \sum_{t=m+1}^{2m} \xi_{3,t} - \mathbb{E} \big[ \xi_{3,t} | \mathcal{F}_{t-1} \big]  
\end{pmatrix}
\end{align*} 
where we denote with
\begin{align}
\xi_{1,t} &=  \boldsymbol{1} \big\{ \varepsilon_t \leq \boldsymbol{v}^{\top} \mathbb{M}_m^{-1} \boldsymbol{\Lambda}_{t-1} \big\} - \boldsymbol{1} \big\{ \varepsilon_t \leq 0  \big\} 
\\
\xi_{2,t} &= \left[ \boldsymbol{1} \big\{ \varepsilon_t \leq \boldsymbol{v}^{\top} \mathbb{M}_m^{-1} \boldsymbol{\Lambda}_{t-1} \big\} - \boldsymbol{1} \big\{ \varepsilon_t \leq 0  \big\} \right] \cdot  w_t
\\
\xi_{3,t} &=  \left[ \boldsymbol{1} \big\{ \varepsilon_t \leq \boldsymbol{v}^{\top} \mathbb{M}_m^{-1} \boldsymbol{\Lambda}_{t-1} \big\} - \boldsymbol{1} \big\{ \varepsilon_t \leq 0  \big\} \right] \cdot  \widetilde{w}_t 
\end{align}
\end{proof}
Notice that the following invariance principles hold:
\begin{align}
\frac{1}{ \sqrt{m} } x_{ \floor{2m r} } \overset{d}{\to} J_c(r) = \int_0^r e^{ (r - s) } dW_u(s)     
\end{align}
\begin{align}
\frac{1}{m} \sum_{t=1}^m \frac{ x_{t-1} }{ \sqrt{m} } \overset{d}{\to} \int_0^1 J_c  \left( \frac{r}{2}  \right) dr    
\end{align}

\newpage

\subsubsection{Conditional Heteroscedasticity Assumptions}

Consider the following formulation of the predictive regression model
\begin{align}
y_t &= \alpha + \beta x_{t-1} + \varepsilon_{y,t}
\\
x_t &= \mu ( 1 - \rho ) + \rho x_{t-1} + \varepsilon_{x,t}
\\
\varepsilon_{y,t} &= \vartheta \varepsilon_{x,t} + \varepsilon_{y.x,t}
\end{align}
Two major problems stem from the observed behaviour of the predictors. The first problem is if $\vartheta \neq 0$, such that $\varepsilon_{y,t}$ and  $\varepsilon_{x,t}$ are correlated, which raises the issue of endogeneity bias. The second problem is the presence of persistence predictors. In other words, although according to theory many of the predictors used should be stationary, empirically most predictors are slowly mean-revering, with a near unit root behaviour (see, \cite{westerlund2015testing}). Moreover, motivated by the possible presence of volatility clustering in stock returns, suitable modifications to the statistical methodology should be employed in order to account for heteroscedasticity similarly to how the degree of endogeneity is modeled to improve the power performance of test statistics. Specifically, modelling conditional heteroscedasticity using a suitable econometric specification, deals with the noisiness of the regression error $( \varepsilon_{y,t} )$, thereby capturing accurately the signal coming from $x_t$, and thus leading to correct inference when returns are truly predictable. 

In particular, a feasible quasi-GLE (FQGLS) $t-$test can be proposed that is robust to misspecification of the conditional variance, which is asymptotically equivalent to GLS if the model is correct. In other words, it can be shown (both analytically and using simulations), that the information contained in the heteroscedasticity is useful, and that the power of the FQGLS test can be increased above that achievable by existing LS-based tests that do not make use of this information. Consider below modelling conditional heteroscedasticity
\begin{align}
\mathbb{E} \big[ \varepsilon^2_{y,t} | \mathcal{F}_{t-1} \big] &= \vartheta^2  \mathbb{E} \big[ \varepsilon_{x,t}^2  | \mathcal{F}_{t-1} \big]  + \mathbb{E} \big[  \varepsilon_{y.x,t}^2 | \mathcal{F}_{t-1} \big] 
\end{align}
which implies that $\sigma^2_{y,t} = \vartheta^2 \sigma^2_{x,t} + \sigma^2_{y.x,t}$. Notice that the FQGLS estimator relies on having an estimator $\hat{s}^2_{y,t}$ of $\sigma^2_{y,t}$. In particular, $\hat{s}^2_{y,t}$ can be the fitted value from a parametric specification of the conditional heteroscedasticity, such as an ARCH model. 
\begin{example}
A commonly used multivariate conditional heteroscedasticity model is the constant-correlation ARCH specification proposed by \cite{bollerslev1990modelling}, which implies that
\begin{align}
\Sigma_{\varepsilon, t} := \mathbb{E} \big[ \varepsilon_t \varepsilon_t^{\prime} \big]   = 
\begin{bmatrix}
\sigma_{x,t}^{2} & \sigma_{xy,t}
\\
\sigma_{xy,t} & \sigma_{y,t}^{2}
\end{bmatrix}
\equiv  
D \Theta_t D^{\prime}
=
\begin{bmatrix}
\sigma_{x,t}^{2}   & \vartheta \sigma_{x,t}^{2} 
\\
\vartheta \sigma_{x,t}^{2} &  \sigma_{y.x,t}^{2} + \vartheta^{2} \sigma_{x,t}^{2}
\end{bmatrix},
\end{align}
where $\varepsilon_t = ( \varepsilon_{x,t}, \varepsilon_{y.x,t} )$,  $D = \begin{bmatrix}
1 & 0
\\
\vartheta & 1
\end{bmatrix}$ and $\Theta_t = \mathsf{diag} \left\{ \sigma_{x,t}^{2}, \sigma_{y.x,t}^{2} \right\}$. 
\end{example}


\newpage

\section{Monte Carlo Simulation Study}
\label{Section3}

The numerical studies of our paper include both a Monte Carlo simulation experiments as well as an empirical application to validate the performance of the proposed statistical methodologies on a real data set. In this Section, we examine the finite sample performance of the proposed estimators and predictability tests via an extensive Monte Carlo simulation study. We consider the finite sample size and power performance of the proposed QR method. 

\subsection{Experimental Design}

The data generating mechanism is given by the following expression: 
\begin{align}
y_t &= \alpha ( \tau ) + \beta (\tau ) x_{t-1}  + u_t ( \tau)
\\
x_t &= \mu + \rho x_{t-1} + v_t, \ \ \rho = \left(  1 - \frac{c}{n} \right) 
\end{align}

\subsection{Empirical size and power analysis of predictability tests}

Before illustrating insights from Monte Carlo simulation study it is useful to discuss main findings presented in the literature. In particular, from Table 1 in \cite{yang2021unified} we summarize the following observations. Under the presence of a non-zero augmented variable for large samples the IVX estimator for testing the null hypothesis $H_0: \beta = 0$ has an excellent empirical size comparable with the empirical likelihood approach given a zero model intercept in the autoregressive equation. On the other hand, under the presence of a non-zero model intercept the size of the IVX-based test statistic detiorates while the corresponding test statistic based on the empirical likelihood approach seems to preserve a good control of size in both finite and large samples. Furthermore, \cite{hosseinkouchack2021finite} propose finite-sample size control for the IVX instrumentation by considering recursive demeaning as a possible correction. Specifically, by using a recursively demeaned regressor and forward demeaning for the regressand implies that the recursively demeaned regressor and the forward demeaned distrurbance are orthogonal irrespective of the correlation between the innovation sequences $u_t$ and $v_t$, thereby ensuring robustness to the degree of endogeneity regardless of the presence of model intercepts in the two functional forms.

\subsubsection{Weighted Bootstrap}

Since the random weighted bootstrap method neither infers the heteroscedastic errors nor generates samples from the fitted model via resampling residuals like residual-based bootstrap methods, the developed test is robust against conditional heteroscedasticity and computational convenient (see,  \cite{liu2023unified} and \cite{hong2023unifying}).

\section{Empirical Application}
\label{Section4}

The stock return predictability is a major puzzle in financial economics. The literature goes back several decades; we briefly highlight important studies. The seminal paper of \cite{perron1989great} discuss the unit root hypothesis around periods of financial turbulence. \cite{dejong1991temporal} study the debate whether divided are trend-stationary or integrated processes while \cite{stein1991stock} propose a framework for stock price distributions and stochastic volatility. We focus on examining the presence of predictability for monthly US stock market excess returns over the period 1990-2019 using the set of variables considered in \cite{welch2008comprehensive}\footnote{The dataset can be retrieved from Amit Goyal's website at \url{http://www.hec.unil.ch/agoyal/}. Detailed descriptions of variables can be found in the Online Appendix of \cite{welch2008comprehensive}. } which capture economic and financial conditions for the US economy. Following the common practise in the predictability literature, we use value weighted returns for the  with monthly sampling frequency from Jan-1990 to Dec-2020. Therefore, we employ the predictive regression model which is an econometric environment that allows for the two innovations of the system to be correlated a more realistic assumption when modelling the presence of cointegrating dynamics between stock returns and macroeconomic and/or financial variables in the economy.

\section{Conclusion}
\label{Section5}

The use of optimization techniques such as the empirical likelihood estimation procedure or the profile likelihood approach can be used for estimating predicting regression models and are found to be robust to unknown degree of persistence is a similar spirit as when the IVX instrumentation is applied. Although these two estimation methodologies work differently there is a common denominator: regardless of the persistence properties of regressors uniform inference can be applied which provides a solution to the challenges that other methodologies in the literature have (such as Bonferonni corrections and the alike). Both procedures are practical in empirical applications and it is up to the econometrician to decide which methodology is preferable especially when developing asymptotic theory results for constructing test statistics. 

In this article we develop a framework for unified statistical inference for dynamic quantile predictive regression models regardless of zero on nonzero intercept, without any prior information on the unknown degree of persistence in modeling possibly nonstationary predictors. Our test statistics are constructed using weighted quantile regression estimation which preserves the asymptotic mixed-Gaussianity (AMG) property of the estimators. Consequently, we can obtain a unified predictability test whose statistic follows a Chi-square distribution. Although the empirical likelihood approach is not always a preferable approach by econometricians it does provide valid uniform inference methodology. On the other hand the IVX filtration is also a robust uniform methodology while allows to develop and extend relevant econometric theory results in various directions.

\newpage 

\paragraph{Conflicts of interest}

The author declares that there are no known conflicts of interest.



\paragraph{ORCID}

Dr. Christis Katsouris \orcidA{} \begin{small} \url{http://www.orcid.org/0000-0002-8869-6641} \end{small}

\bibliographystyle{apalike}
\bibliography{myreferences1}

\end{document}